\documentclass[11pt,letterpaper]{article}
\usepackage[T1]{fontenc}
\usepackage[utf8]{inputenc}
\usepackage[margin=1in]{geometry}
\usepackage{lmodern}
\usepackage{microtype}
\usepackage{amsmath,amssymb,amsthm,mathtools}
\usepackage{enumitem}
\usepackage{booktabs,tabularx,array}
\usepackage[numbers,square,sort&compress]{natbib}
\usepackage[hidelinks]{hyperref}
\usepackage[capitalize,noabbrev]{cleveref}
\usepackage{url}
\usepackage{tikz}
\usepackage{float}
\setlist[itemize]{topsep=3pt,itemsep=2pt,leftmargin=1.55em}
\setlist[enumerate]{topsep=3pt,itemsep=2pt,leftmargin=1.65em}
\newtheorem{theorem}{Theorem}[section]
\newtheorem{lemma}[theorem]{Lemma}

\newtheorem{corollary}[theorem]{Corollary}

\theoremstyle{definition}

\newcommand{\OPT}{\operatorname{OPT}}

\newcommand{\BU}{\mathrm{BU}}
\newcommand{\OPE}{\mathrm{OPE}}
\newcommand{\DIR}{\mathrm{DIR}}
\newcommand{\cov}{\operatorname{cov}}

\newcommand{\poly}{\operatorname{poly}}
\newcommand{\Topt}{T^\star}
\newcommand{\Uopt}{U^\star}
\hypersetup{pdftitle={The Algorithmic Limits of Tokenisation: Approximation, Exact Algorithms, and Unary Merges},
pdfauthor={Andrei Robert Ion, Rares Mateiu, Adrian Miclaus, Alexandru Popa}}
\title{How Hard Is Tokenisation? Algorithms and Complexity Beyond NP-Hardness}
\author{%
Andrei Robert Ion\textsuperscript{1}\qquad
Rares Mateiu\textsuperscript{2}\\[0.3em]
Adrian Miclaus\textsuperscript{2}\qquad
Alexandru Popa\textsuperscript{2}\\[0.75em]
\normalsize\textsuperscript{1}\'Ecole Polytechnique, France\\[0.15em]
\normalsize\textsuperscript{2}University of Bucharest, Romania
}
\date{}
\begin{document}
\hypersetup{pageanchor=false}
\begin{titlepage}
\maketitle
\thispagestyle{empty}
\vspace{0.65em}
\begin{abstract}
Given a collection of strings and a vocabulary budget, how well can we
minimize its total token count, and how does the answer change with the
tokenisation model? For binary direct tokenisation, previous work proved
APX-hardness but left constant-factor approximability unresolved
\citep{kastreva2026}. We rule out every polynomial-time
$M^{1-\varepsilon}$ approximation, for fixed $\varepsilon>0$, unless
$\mathrm{P}=\mathrm{NP}$, even on distinct equal-length strings with
$K=m-1$. Here $M$ is the total input length and $m$ is the number of
input strings. We complement this lower bound with a deterministic
polynomial-time $O(M/\log M)$ approximation algorithm and stronger
bounds for selected vocabulary budgets. For the different objective of
maximizing compression savings, GreedTok was shown to attain a constant
fraction of the optimum on unrestricted alphabets \citep{baliga2026};
we establish a deterministic EPTAS for direct savings on every fixed
alphabet and analyze greedy selection by true marginal gains.
For exact optimization, we give a binary direct algorithm running in
$2^{O(n\log\log n/\log n)}\poly(I)$ time, where $n$ is the total length
of the distinct input strings and $I$ is the input encoding length.
Finally, we resolve two unary merge-model hardness questions left by
the bounded-alphabet study \citep{kastreva2026}: standard unary bottom-up
tokenisation is APX-hard, and unary optimal pair encoding is strongly
NP-complete, strengthening the previously known weak hardness.
\end{abstract}
\vfill
\end{titlepage}
\hypersetup{pageanchor=true}
\pagenumbering{arabic}

\section{Introduction}\label{sec:introduction}
\subsection{Motivation}
Tokenisation replaces a string by a sequence of words from a chosen vocabulary.
Shorter sequences can reduce storage needs and the amount of work performed by
sequence models. In many systems, the vocabulary is built by a heuristic.
Subword vocabularies are central to practical systems based on byte-pair
encoding and related tokenisers~\citep{sennrich2016,kudo2018}.
A basic theoretical question is whether we can find a small vocabulary that
minimizes the total number of tokens.

This question depends on the rules used to construct a vocabulary and parse
its strings. In direct tokenisation, a vocabulary is chosen first and every
string is then segmented optimally. In bottom-up tokenisation, a sequence of
merge rules changes the current parse. The resulting parse need not be the
shortest parse over the words created by those rules.
We study both models and show how their computational properties differ.

\subsection{Previous and related work}
\citet{whittington2025} proved NP-completeness of direct and bottom-up
tokenisation. \citet{kastreva2026} established APX-hardness for binary
instances and hardness for unary direct tokenisation. These results leave
open the possibility of approximation algorithms for the token-count
objective, even over a binary alphabet.

\citet{lim2025} formulated partition-cover tokenization and introduced
GreedTok and the maximum-coverage relaxation GreedWMC.
\citet{baliga2026} subsequently proved that GreedTok achieves at least
$0.2752$ of the optimum compression savings and established an
inapproximability bound for savings over unrestricted alphabets.
Their partition-cover savings objective is the same as $U_D(V)$ when
segmentations are optimal. Our fixed-alphabet approximation scheme concerns
this savings objective, whereas our near-linear inapproximability and
$O(M/\log M)$ approximation concern the different multiplicative objective
of final token count. The constant-factor guarantees for savings do not imply
constant-factor guarantees for token count.

\citet{kozma2026} analysed the compression utility of byte-pair encoding and
optimal pair encoding. \citet{tempus2026} developed ConvexTok through a
linear-programming formulation, and \citet{optfsst2026} studied static
symbol-table selection with optimal parsing for a fixed table. We focus on
worst-case guarantees for the stated combinatorial tokenisation models.

\subsection{Informal problem definition}
We are given a list of strings and a budget $K$.
In the direct model, the single characters are available for free and we may
select at most $K$ additional words. Every input is split into the fewest
possible vocabulary words. We minimize the total token count.
We also study the equivalent exact objective of maximizing the number of
characters saved by the tokenisation; the two objectives need not have the
same approximation guarantees.

For bottom-up tokenisation, we select at most $K$ ordered pair-merge rules.
Each rule is applied from left to right to all nonoverlapping occurrences in
the current parses, and the strings are not resegmented afterward.
Optimal pair encoding instead uses the merge rules to construct a vocabulary
and then permits optimal segmentation with that vocabulary.

\subsection{Our results}
Our binary direct token-count lower bound strengthens the binary
APX-hardness of Kastreva et al.~\citep{kastreva2026}: no deterministic
polynomial-time $M^{1-\varepsilon}$ approximation is possible for any
fixed $\varepsilon>0$, unless $\mathrm P=\mathrm{NP}$. In particular,
the constant-factor approximation question has a negative answer.
We complement this with a polynomial-time $O(M/\log M)$ algorithm and
budget-sensitive improvements.

For compression savings, Baliga et al.~\citep{baliga2026} proved a
constant-factor guarantee for GreedTok and APX-hardness over unrestricted
alphabets. In contrast, our direct-savings EPTAS applies when the alphabet
is fixed. We also analyze a separate algorithm that selects token types
using their true marginal savings under optimal segmentation.

Our exact algorithms use the structure of token occurrences in an optimal
solution. They include a $2^{O(n\log\log n/\log n)}\poly(I)$ algorithm
for binary direct tokenisation, together with a faster unary algorithm and
fixed-parameter results for several measures of input sharing.
For merge-based tokenisation, we establish APX-hardness of standard unary
bottom-up tokenisation, for which the earlier bounded-alphabet work had
no hardness result, and strong NP-completeness of unary optimal pair
encoding, strengthening its earlier weak NP-hardness~\citep{kastreva2026}.

\section{Preliminaries and Tokenisation Models}\label{sec:preliminaries}
Let $D=(s_1,\ldots,s_m)$ be an explicitly represented list of independent nonempty strings over a finite alphabet $\Sigma$. Repetitions are allowed; $M=\sum_i|s_i|$ is the total unweighted input length. The mandatory vocabulary contains the $|\Sigma|$ single characters. A direct solution is a set $V$ of at most $K$ additional words, each of length at least two. Let $\tau(s_i,V)$ denote the minimum number of vocabulary words whose concatenation is $s_i$. We write $T_D(V)=\sum_i\tau(s_i,V)$ and $U_D(V)=M-T_D(V)$, with $\Topt=\min_{|V|\le K}T_D(V)$ and $\Uopt=\max_{|V|\le K}U_D(V)$. Thus $\Uopt=M-\Topt$, but a multiplicative approximation for one objective need not approximate the other multiplicatively.

A bottom-up solution consists of an ordered sequence of at most $K$ rules $(x,y)\mapsto xy$. Every rule is applied exhaustively from left to right on the \emph{current token parses} of each input; merges never cross input boundaries. The final token count is that of the actual resulting parses. Unary tokens can equivalently be represented by their positive integer lengths, but input strings themselves are written explicitly. We use $I$ for the full encoding length and $n$ for the total length after duplicate input strings are removed, when appropriate. All logarithms in asymptotic bounds are base two unless stated otherwise.

\subsection{DIRECT, bottom-up, and optimal pair encoding}
For a dataset of unary strings, identify the token $a^x$ with its positive integer length $x$. The three models share the mandatory character token and a budget of at most $K$ added token types or merge rules. DIRECT chooses up to $K$ literal types and uses a minimum-token parse. In standard bottom-up tokenisation, at most $K$ ordered pair-merge rules are executed exhaustively from left to right on the current parses, without resegmentation. In optimal pair encoding (OPE), binary concatenation rules generate a vocabulary and the strings are then segmented optimally using that vocabulary~\citep{kastreva2026}.

Write $L^*_{\DIR}(D,K)$, $L^*_{\OPE}(D,K)$ and $L^*_{\BU}(D,K)$ for the three optimal token counts. The models satisfy
\begin{equation}\label{unaryope:eq:sandwich}
L^*_{\DIR}(D,K)\le L^*_{\OPE}(D,K)\le L^*_{\BU}(D,K).
\end{equation} These inequalities do not by themselves transfer hardness; the reductions below analyze the corresponding models separately. In all cases, the input strings are explicit and merges cannot cross string boundaries.
\section{Polynomial-Time Approximation for Binary DIRECT}\label{sec:apxcount}
The first family of algorithms respects the exact optional vocabulary budget and includes both a uniform $O(M/\log M)$ guarantee and refinements depending on $K$ and on the vocabulary deficit.

\subsection{Notation and elementary bounds}
An instance is an explicit list of binary strings, a nonnegative integer budget $K$, and mandatory vocabulary $\{0,1\}$. One may add at most $K$ arbitrary binary words and must parse every listed string using the resulting vocabulary. The objective is the sum of token counts. Empty strings may be deleted. Let $M$ be the total number of input characters, counting repeated copies; let $m$ be the number of nonempty strings counting multiplicity; and let $q$ be the number of distinct nonempty strings. Thus $q\leq m\leq M$, and every feasible solution has cost at least $m$. All logarithms are base two; set
 $\displaystyle \ell=\lceil\log_2(M+1)\rceil,\qquad g=\max\{1,\lceil\log_2\ell\rceil\}.$ 
The bounds below concern $M>0$; an empty input is handled exactly.

\subsection{Budget-dependent bounds}
Selecting the $K$ largest whole-input gains $f_s(|s|-1)$ yields cost at most $M-(K/q)(M-m)$. For $1\le K<q$, division by $\OPT\ge m\ge q$ and $(q-2K)^2\ge0$ give ratio at most $1+M/(4K)$. For distinct common-length $a$ inputs with $K=q-1$, its exhibited cost is $q-1+a$, hence ratio at most $1-1/q+M/q^2$.
For $K\ge4$, set $b=\lfloor\log_2(K+4)\rfloor-1$. All binary words of lengths two through $b$ number $2^{b+1}-4\le K$. Block parsing costs at most $M/b+m(b-1)/b$, so the ratio is at most $M/[b(K+1)]+(b-1)/b$ when $K<q$. Taking the better candidate gives $O(1+M/(K\log K))$. Large budgets use the exact whole-input shortcut. These bounds concern the same exact optional budget and are useful independently of the uniform grid bound.
\subsection{General deterministic bound}
\begin{theorem}
There is a deterministic algorithm for binary DIRECT whose approximation ratio is at most $33M/\ell$. Its running time is $O(\text{input-reading cost}+M^2\log^2(M+2))$ word-RAM operations, or $O(\text{input-reading cost}+M^2\log^3(M+2))$ under the conservative bit-operation bound. The polynomial exponent is independent of $K$.
\end{theorem}
\begin{proof}
The proof combines a finite grid search for a small budget with a universal short-block vocabulary for a larger budget.

First handle $K=0$ exactly with the mandatory characters. If $K\geq q$, choosing all distinct input strings of length at least two solves the instance exactly with cost $m$. Every remaining case has $1\leq K<q\leq m$, and $\mathrm{OPT}\geq m\geq q\geq K+1$.

Put $\ell=\lceil\log_2(M+1)\rceil$. If $m\geq\ell$, the character solution costs $M$, so its ratio is at most $M/m\leq M/\ell$. If $1\leq K\leq3$ and $\ell\leq32(K+1)$, return the character dictionary; its ratio is at most $M/(K+1)\leq32M/\ell$.

For the remaining cases set $g=\lceil\log_2\ell\rceil$ and $k_0=\lfloor\ell/(4g)\rfloor$. If $1\leq K\leq3$ escaped the character branch, then $\ell>32(K+1)$, which implies $K\leq k_0$. Indeed, $\ell\geq65$ and $g\geq7$. If $g=7$, then $\ell\leq128$, so $K=3$ is impossible and $\ell/(4g)\geq65/28>2\geq K$. If $g\geq8$, then $\ell>2^{g-1}$, so $\ell/(4g)>2^{g-1}/(4g)\geq4>K$; the middle expression is four at $g=8$ and increases with $g$. Thus a small positive budget not handled by the character branch reaches the grid branch. In particular, the universal-block branch below is used only when $K\geq4$. The cutoff $k_0$ may be zero.

We use a grid rounding fact. In each input string, place grid points at positions $0,\Delta,2\Delta,\ldots$, together with the final endpoint, removing duplicates. Let $C$ be the set of distinct substrings of length at least two between pairs of grid points in the same string. For any dictionary of at most $K$ optional words and any parse of cost $T$, there is a dictionary of at most $K$ words from $C$ with cost at most $(2\Delta+1)T$. To see this, choose one used occurrence for each optional word type $w$. If its closed interval contains at least two grid points, let the first and last give a core $c(w)$. Then $w=p c(w) r$ with $|p|,|r|<\Delta$. This is a fixed factorization of the word, so it works for every occurrence of that type, even at other grid phases. Replace each token by its two character fringes and the one core token (if the core has length one it is mandatory). If fewer than two grid points lie in the interval, its length is less than $2\Delta$, and characters suffice. Each original token is replaced by at most $2\Delta+1$ tokens, and each old optional type creates at most one optional core.

If $K\leq k_0$, take $\Delta=\lceil M/\ell\rceil$. Whenever this grid branch is reached, either $K\geq4$ and the earlier cases imply $m<\ell$ with $m\geq5$, or $K\leq3$ and the dichotomy above gives $\ell\geq65$. Thus $g\geq3$. The number $P$ of grid points over all listed strings is at most $M/\Delta+2m\leq\ell+2m<3\ell$; hence $Q=|C|\leq P^2<9\ell^2$. Enumerate all subsets of $C$ of size at most $K$, and return one of minimum segmentation cost. Dummy-padded sorted tuples bound the number of candidates by $(Q+1)^K$. Since $g\geq3$,
 $\displaystyle (Q+1)^K\leq 2^{4gK}\leq2^\ell\leq2M.$ 
Rounding an optimal solution therefore shows that the returned ratio is at most
 $\displaystyle 2\Delta+1\leq2M/\ell+3\leq5M/\ell.$ 

If $K>k_0$, choose $b=\lfloor\log_2(K+4)\rfloor-1$. Since $K\geq4$, $b\geq2$, and the dictionary containing every binary word of lengths $2,\ldots,b$ has $2^{b+1}-4\leq K$ words. Parsing in blocks of length at most $b$ costs at most $M/b+m$, giving ratio at most $M/(bm)+1$. One checks $b\geq g/8$: for $g\leq16$ this follows from $b\geq2$; for $g\geq16$, $K>\ell/(4g)$ implies $b>g-5-\log_2g\geq g/2-5\geq g/8$. Also $m\geq q>K>\ell/(4g)$, so $bm>\ell/32$. The ratio is therefore less than $32M/\ell+1\leq33M/\ell$.

Every branch uses the original budget. The grid branch evaluates at most $2M$ dictionaries, each by polynomial-time segmentation DP; the character and exact branches are linear or polynomial, and the universal dictionary is polynomial in size. This establishes the theorem.
\end{proof}

\subsection{Vocabulary-deficit bounds}
\begin{theorem}
Let $d=q-K>0$. There is a deterministic algorithm for binary DIRECT with approximation ratio at most
 $\displaystyle 3+\frac{64dMg^2}{\ell^2}.$ 
It respects the exact budget and runs in $O(\text{input-reading cost}+M^2\log^2(M+2))$ word-RAM operations, or $O(\text{input-reading cost}+M^2\log^3(M+2))$ under the conservative bit-operation bound. Here $q$ counts distinct nonempty strings; repetitions are allowed and are counted in $M,m$. No equal-length assumption is needed.
\end{theorem}
\begin{proof}
The empty instance, $K=0$, and $K\geq q$ are exact. In the remaining cases $1\leq K<q\leq m$, so $d=q-K\geq1$, $M\geq2$, and $\ell\geq2$. Use $g=\max\{1,\lceil\log_2\ell\rceil\}$.

For a distinct string $s$ with multiplicity $f_s$, define its whole-word gain $a_s=f_s(|s|-1)\geq0$. Selecting the $K$ types with largest gains leaves the $d$ smallest gains omitted. Their sum is at most $d/q$ times the total gain, namely $d(M-m)/q$. The resulting whole-word/character parse costs at most $m+d(M-m)/q$. Since $\mathrm{OPT}\geq m\geq q$, its ratio is at most
 $\displaystyle 1+\frac{d(M-m)}{qm}\leq1+\frac{dM}{q^2}.$ 
A length-one type has zero gain and already parses with a mandatory character, so formal selection of such a type does not use more than $K$ entries.

Set $k_1=\lfloor\ell/(8g)\rfloor$ and $\Delta=\lceil M/\ell^2\rceil$. Use these three branches after the exact cases.
\begin{enumerate}
\item If $d\geq\ell/(8g)$, return the mandatory characters. Then $q\geq d$, and
 $\displaystyle \frac{\mathrm{ALG}}{\mathrm{OPT}}\leq\frac Mq\leq\frac Md\leq\frac{64dMg^2}{\ell^2}.$ 
\item If $d<\ell/(8g)$ and $K>k_1$, use the weighted whole-word solution above. Since $q>K>\ell/(8g)$, its ratio is less than $1+64dMg^2/\ell^2$.
\item Otherwise $d<\ell/(8g)$ and $K\leq k_1\leq\ell/(8g)$. Thus $q=K+d<\ell/(4g)\leq\ell$. Use grid rounding on distinct input strings with spacing $\Delta$. The number of grid points satisfies
 $\displaystyle P\leq M/\Delta+2q\leq\ell^2+2\ell\leq3\ell^2,$ 
so the raw and distinct candidate counts are at most $9\ell^4$. Because $\ell\leq2^g$ and $g\geq1$, $Q+1\leq9\cdot2^{4g}+1\leq2^{8g}$. The number of dictionaries of size at most $K$ is at most $2^{8gK}\leq2^\ell\leq2M$. Fixed-core rounding gives ratio
 $\displaystyle 2\Delta+1\leq3+2M/\ell^2\leq3+64dMg^2/\ell^2.$ 
\end{enumerate}
These cases cover every input and use at most $K$ optional words. In the grid case, $q<\ell$, and the number of dictionaries is within the stated running time. Repeated copies share the same candidate words and do not increase this count. This proves the bound.
\end{proof}
For $d=1$, this recovers the previous $K=q-1$ result. More generally, for each fixed deficit $d$, the guarantee is $O(dM(\log\log M)^2/(\log M)^2)$. Taking the better of this algorithm and the general grid algorithm gives ratio at most
 $\displaystyle \min\left\{\frac{33M}{\ell},\;3+\frac{64dMg^2}{\ell^2}\right\}.$ 

\subsection{Uniform implementation bounds}
The runtime bounds include reading the encoded corpus/list and budget. If arbitrarily many empty strings are permitted, their list metadata is not bounded by $M$. With only nonempty listed strings, list metadata is $O(M)$.

For the general theorem, the branch order is: discard empty strings; solve the empty corpus, $K=0$, and $K\geq q$ exactly; use characters when $m\geq\ell$; use characters also when $1\leq K\leq3$ and $\ell\leq32(K+1)$; then, if $K\leq k_0=\lfloor\ell/(4g)\rfloor$, enumerate grid dictionaries with $\Delta=\lceil M/\ell\rceil$; otherwise use all binary words of lengths $2,\ldots,\lfloor\log_2(K+4)\rfloor-1$. The small-budget dichotomy in the proof ensures any $K\leq3$ escaping the character branch satisfies the grid cutoff. Thus the universal-block branch is used only when $K\geq4$. There is no exact arbitrary-substring enumeration.

In the general grid branch, there are $R=O(\ell^2)$ raw grid intervals and $Q=O(\ell^2)$ distinct candidate words, each of length at most $M$. Copying and trie-deduplicating the raw words costs $O(RM)=O(M\ell^2)$. Find occurrences by comparing each distinct candidate against every possible corpus start, stopping at its own string boundary; this takes $O(QM^2)$ character comparisons. The resulting position DAG has $O(M(Q+1))$ optional and mandatory edges. For each selected dictionary, a forward shortest-path DP skips optional edges whose candidate identifiers were not selected and examines $O(M(Q+1))$ edges. At most $2M$ dictionaries are evaluated, for total $O(M^2(Q+1))=O(M^2\ell^2)$ DP work. Candidate masks, storage of the best set, and one final predecessor pass fit within this bound. Hence the word-RAM running time is $O(\text{input-reading cost}+M^2\log^2(M+2))$, with conservative bit-operation bound $O(\text{input-reading cost}+M^2\log^3(M+2))$.

For the vocabulary-deficit theorem, the grid branch applies whenever the third-branch conditions hold, in particular $K\leq\lfloor\ell/(8g)\rfloor$; $K=0$ is solved exactly. No separate exact search for $K\leq3$ is needed. In this branch, $Q\leq9\ell^4$ and $Q+1\leq2^{8g}$, so at most $2^{8gK}\leq2^\ell\leq2M$ dictionaries are evaluated.

For the sharper grid-branch runtime, let $M_{\mathrm{dist}}\leq M$ be the total length of distinct input strings. Differences between regular grid points are multiples of $\Delta=\lceil M/\ell^2\rceil$, contributing at most $\lfloor M_{\mathrm{dist}}/\Delta\rfloor$ candidate lengths. Each string has at most one irregular final endpoint; differences from it to regular grid points add at most $\lfloor |s|/\Delta\rfloor+1$ lengths. Summing over distinct strings gives
 $\displaystyle H_{\mathrm{len}}\leq2\lfloor M_{\mathrm{dist}}/\Delta\rfloor+q=O(\ell^2).$ 
At a fixed corpus position and fixed length, the binary substring is unique. Since candidates are deduplicated, at most one candidate of that length can match there. Thus the optional occurrence graph has $E=O(M\ell^2)$ edges.

Build a binary trie from the raw candidate words, assigning one identifier per terminal. Aho--Corasick matching finds every candidate occurrence in $O(M+E)$ time after $O(RM)=O(M\ell^4)$ trie construction. The alphabet is binary, so transitions have constant cost. The original construction has failure/output structures whose preprocessing is linear in total pattern length, and its scan makes fewer than $2n$ state transitions on text of length $n$; emitting matches costs $O(E)$. For each of at most $2M$ dictionaries, forward DP over $O(M+E)$ position-DAG edges costs $O(M+E)$, totaling $O(M^2\ell^2)$. Mask work $O(MQ)=O(M\ell^4)$ and trie construction also fit this total because $\ell^2=O(M)$. Memory is $O(M\ell^4)$ for the trie and $O(M\ell^2)$ for the occurrence graph. The whole-word branch aggregates equal strings with a binary trie and sorts the $q$ weighted gains in $O(q\log(q+2))$ time, within the stated bound.

Aho--Corasick matching finds all candidate occurrences in time linear in the total candidate length, the input length, and the number of matches \citep{ahocorasick1975}. Each reported match becomes one edge of the segmentation graph, as counted above.

\section{Near-Linear Inapproximability}\label{sec:hardcount}
We prove hardness by encoding an instance of finite-word combinatorial rank as binary strings. We then show how a short tokenisation yields a solution to the original instance. The same construction gives the equal-length restriction.

\subsection{Problem statement and result}

An instance consists of a finite explicit list \(S=(s_1,\ldots,s_m)\) of
nonempty strings over \(\{0,1\}\), and a nonnegative integer \(K\). The
mandatory vocabulary is \(\{0,1\}\). A feasible optional vocabulary \(V\) is
any set of at most \(K\) additional nonempty binary words; candidates need not
be specified in the input. Its cost is
 $\displaystyle \operatorname{cost}_S(V)= \sum_{i=1}^m\min\{r:s_i=w_1\cdots w_r,\ w_j\in\{0,1\}\cup V\}.$ 
Binary DIRECT minimizes this cost. An optional word unused in every parse
can be deleted, so restricting attention to input substrings would not change
the optimum. Mandatory character tokens do not count against \(K\). Write
 $\displaystyle M=\sum_{i=1}^m |s_i|$ 
for the total explicit number of binary characters.

The source problem is finite-word combinatorial rank. Given a finite set
\(X\subseteq\Sigma^+\) of explicit nonempty words over a finite,
input-dependent alphabet and an integer \(K\), decide whether
 $\displaystyle r(X):=\min\{|Y|:X\subseteq Y^*\}\le K,$ 
where \(Y\subseteq\Sigma^+\) is an arbitrary set of nonempty words. No source
alphabet generators are supplied for free, and there is no bound on the
number of factors in a factorization. N\'eraud proved this bounded-rank
decision problem NP-complete \citep[Theorem 4.5]{neraud1990}. This is
combinatorial rank, distinct from free/code rank.

\begin{theorem}For every fixed $\epsilon\in(0,1)$, a deterministic polynomial-time $M^{1-\epsilon}$ approximation for binary DIRECT would imply $\mathrm P=\mathrm{NP}$. This remains true jointly for distinct equal-length inputs and budget $K=m-1$. These exclusions also imply the corresponding fixed-constant-factor exclusions.\end{theorem}
\subsection{Synchronizing binary encodings}

Let \(X=(x_1,\ldots,x_m)\) be a source instance, \(N=\sum_i|x_i|\), and let
\(\Sigma\) be the letters appearing in \(X\), so \(|\Sigma|\le N\). Choose a
positive block length \(L\), and set
 $\displaystyle M=NL,\qquad h=\left\lceil24\log_2(2M)\right\rceil.$ 
The applications below have \(N\ge2\), \(h<L\), and polynomially bounded
\(L\). Encode each \(a\in\Sigma\) by \(c(a)\in\{0,1\}^L\), extending \(c\) to
words by concatenation. The encoded corpus length is \(M\).

For a length-\(h\) window in an encoded input, its phase is its starting
position modulo \(L\). Since \(h<L\), it meets at most two source-letter
blocks. Its dominant source letter is the letter of the block containing at
least half of its positions, with ties resolved in favor of the left block.
The selected side is determined by the phase alone.

\begin{lemma}[Synchronizing code]
For every source corpus and every \(L,h\) as above with \(h<L\), there is a
deterministic polynomial-time constructible code \(c:\Sigma\to\{0,1\}^L\)
such that any two length-\(h\) windows occurring in the encoded corpus with
identical binary contents have the same phase and dominant source letter.
\end{lemma}

\begin{proof}
Initially choose independently and uniformly one bit for each
\((a,j)\in\Sigma\times\{0,\ldots,L-1\}\). All occurrences of the same source
letter use the same block. A window starting at phase \(p\) reads \(h\)
source-bit identifiers. These are distinct: if the window crosses a block
boundary, its left suffix and right prefix have total length \(h<L\), so
even equal adjacent source letters use disjoint offsets.

Consider two windows whose contexts (phase, dominant source letter) differ.
Equality of their contents imposes one equality constraint per position.
Remove comparisons of an identifier with itself and form the resulting
multigraph on source-bit variables. Each window contains distinct variables,
so every vertex has degree at most two. Each nontrivial component is a path
or a cycle, including a possible two-edge cycle from parallel edges. If there
are \(q\) non-loop comparisons, equality rank is at least \(q/2\): a path
with \(e\) edges has rank \(e\), while a cycle with \(e\ge2\) edges has rank
\(e-1\ge e/2\). Thus the collision probability is at most \(2^{-q/2}\).

Different phases imply \(q=h\), because every paired identifier has a
different offset. If phases agree but dominant letters differ, the dominant
side is the same in both windows and has at least \(h/2\) positions. At all
these positions the source-letter identifiers differ, so \(q\ge h/2\).
Each bad-context collision therefore has probability at most \(2^{-h/4}\).
There are at most \(M\) occurring windows and fewer than \(M^2\) pairs, so
the expected number of bad collisions is at most
 $\displaystyle M^2 2^{-h/4}\le M^2(2M)^{-6}<1.$ 

We derandomize by conditional expectations. Enumerate bad-context pairs
using their source-bit identifiers, phases, and dominant letters, and
precompute the connected components of each equality graph. Under a partial
bit assignment, an event has conditional probability zero if some component
contains assigned bits of both values. Otherwise, a component with \(u\)
unassigned variables contributes \(2^{-u}\) if it has an assigned bit, and
\(2^{-(u-1)}\) if it has no assigned bit. Multiplication gives the exact
conditional probability. Assign the at most \(|\Sigma|L\le M\) variables
one at a time, each time choosing the bit that minimizes the conditional
expected number of bad events. The expectation never increases; it starts
below one, so the final integer number of bad events is zero.

There are at most \(M^2\) events, each involving at most \(2h\) variables.
Their probabilities are dyadic rationals with denominators dividing
\(2^{2h}\), so exact sums and updates have polynomial bit complexity.
Even recomputing every event probability for every variable is polynomial in
\(M\). This proves the lemma.
\end{proof}

\subsection{Decoding small-cost parses}

Fix any code satisfying the synchronizing property above, with \(1\le h<L\).
The decoding argument uses only this property and \(h<L\); it does not depend
on the probabilistic construction or the specific choice
\(h=\lceil24\log_2(2M)\rceil\). Let \(T\) be the total number of tokens in any
collection of valid parses of the encoded corpus using at most \(K\) optional
binary words beyond \(0,1\).

\begin{lemma}[Marker decoding]
If
\[
 (2h+1)(T+m+1)<L, \tag{1}
\]
then \(r(X)\le K\).
\end{lemma}

\begin{proof}
Include every token boundary in every parse, including both endpoints of each
string. There are \(T+m\) boundary incidences. Each boundary coordinate
\(b\) forbids phases at circular distance at most \(h\) modulo \(L\), at most
\(2h+1\) phases. Also forbid phases within circular distance \(h\) of zero,
so markers stay inside code blocks. By (1), fewer than \(L\) phases are
forbidden. Choose an allowed phase \(t\in\{0,\ldots,L-1\}\).

Place a marker at offset \(t\) in every source-letter code block. Each marker
is more than \(h\) positions from every token boundary and both endpoints of
its source block. Its forward length-\(h\) window therefore lies wholly
inside one token occurrence and one source-letter block. A token containing
a marker is longer than \(h\); neither mandatory one-character token can
contain a marker.

We define one decoded source word per used binary token type. Let \(w\) be a
used optional word with a marker in one occurrence. Its first \(h\) bits are
identical in all occurrences; the synchronizing property implies that every
occurrence starts at the same phase. Since the length is fixed, the relative
marker positions are identical in all occurrences. At each such relative
position, the forward \(h\)-bit window is fixed by \(w\), lies inside one
source block in each occurrence, and has the same phase. Synchronization
therefore gives the same dominant source letter in all occurrences. The
ordered marked letters define a single word \(\operatorname{decode}(w)\).

A used word with no marker is decoded as the empty word. This is consistent:
if any occurrence contains a marker, the common first-\(h\)-window argument
forces every occurrence to have the same phase, and hence the same relative
marker positions. In particular the mandatory tokens \(0\) and \(1\) decode
to the empty word.

Take the nonempty decoded images of all used optional words and remove
duplicates, obtaining \(Y\subseteq\Sigma^+\) with \(|Y|\le K\). Each encoded
source-letter block has exactly one marker, every marker belongs to exactly
one token occurrence, and concatenation preserves marker order. Thus the
decoded images of the tokens in any parse of \(c(x_i)\) concatenate to
\(x_i\). Hence \(X\subseteq Y^*\) and \(r(X)\le K\).
\end{proof}

By contraposition, on a source NO instance, every parse using at most \(K\)
optional words satisfies
\[
 T\ge \frac{L}{2h+1}-m-1. \tag{2}
\]
This is a real-valued lower bound; it follows because (1) must fail.

If $r(X)\le K$, encoding a generating dictionary gives a parse with at most $N$ tokens, since every source factor is nonempty. Thus $\mathrm{OPT}_{\rm bin}\le N$.\par
\subsection{Near-linear inapproximability}

Fix \(0<\varepsilon<1\), and choose the fixed integers
 $\displaystyle d=\left\lceil\frac4\varepsilon\right\rceil,\qquad A=48d+99,\qquad N_0=4A.$ 
We may assume \(N\ge N_0\). If \(N<N_0\), add \(p=N_0-N\) fresh letters as
singleton source strings and replace \(K\) by \(K+p\). Every new singleton
forces its own one-letter generator. No generator containing a fresh letter
can help an old string, so the new rank equals the old rank plus \(p\).
This padding preserves YES/NO and makes the source length \(N_0\). The
padding is constant for the fixed \(\varepsilon\).

For the padded instance choose
 $\displaystyle L=N^d,\qquad M=NL=N^{d+1},\qquad h=\left\lceil24\log_2(2M)\right\rceil.$ 
The output is polynomial in \(N\) since \(d\) is fixed. For \(N\ge2\),
 $\displaystyle 2h+1\le51+48(d+1)\log_2N\le A\log_2N.$ 
Since \(N\ge4A\) and \(\log_2N\le N\),
\[
 \frac{N^2}{A\log_2N}\ge\frac NA\ge4,\qquad
 \frac{L}{2h+1}\ge4N^{d-2}. \tag{6}
\]
Also \(h\le(A/2)\log_2N\le AN/2\le N^2/8<L\). Since \(m\le N\), (2)
implies that on a source NO instance
\[
 \mathrm{OPT}_{\rm bin}\ge4N^{d-2}-N-1>N^{d-2}. \tag{7}
\]
On YES instances, encoding the source generators gives \(\mathrm{OPT}_{\rm bin}\le N\). An
\(M^{1-\varepsilon}\)-approximation returns a feasible solution of cost at
most
\[
 M^{1-\varepsilon}N
 =N^{(d+1)(1-\varepsilon)+1}
 =N^{d+2-\varepsilon(d+1)}
 \le N^{d-2}, \tag{8}
\]
because \(\varepsilon(d+1)\ge4\). Evaluate the returned dictionary's true
cost by dynamic programming and compare with the integer threshold
\(N^{d-2}\). Equations (7)--(8) distinguish YES and NO in deterministic
polynomial time, contradicting NP-completeness unless \(\mathrm{P}=\mathrm{NP}\).
This proves the bound.

The theorem quantifies separately over each fixed \(\varepsilon\). It does
not provide one reduction excluding all sublinear factors simultaneously or
establish hardness of a linear factor. No PCP approximation gap is needed:
the source is an exact rank decision problem, and the target gap comes from
the long encoding and marker decoder.

\subsection{Equal-length inputs with budget one below corpus size}

We prove the same inapproximability bounds when the binary corpus consists of
distinct equal-length strings and \(K=m-1\). The source is N\'eraud's
NP-complete simplifiability problem: given a finite set \(X\) of \(m\)
nonempty words, decide whether \(r(X)\le m-1\). This is the complement of
elementariness in Theorem 5.1 \citep{neraud1990}.

Let \(F=\max_{x\in X}|x|\), choose a fresh letter \(\delta\), and set
 $\displaystyle Z=\{x\delta^{F-|x|}:x\in X\}\cup\{\delta^F\}.$ 
The \(m+1\) words of \(Z\) are distinct and all have length \(F\). They
satisfy
 $\displaystyle r(Z)=r(X)+1.$ 
For the upper bound, adjoin \(\delta\) to a minimum generating set for
\(X\). For the lower bound, any generating set for \(Z\) contains a
nonempty word made only of \(\delta\)'s, since it generates \(\delta^F\).
Erase \(\delta\) from every generator. The images generate \(X\), and at
least one image is empty; after removing empty images and duplicates, at most
\(|Y|-1\) nonempty generators remain. Thus \(r(X)\le r(Z)-1\). Therefore
\(X\) is simplifiable exactly when \(r(Z)\le m\).

The synchronization property makes the letter code injective: if two
appearing letters had the same codeword, their phase-zero length-\(h\)
windows would have identical bits but different dominant letters. Its
concatenation extension therefore maps distinct source words to distinct
binary strings, all of common length \(FL\). Use this corpus as the input to
the same Binary DIRECT reduction with optional-vocabulary budget
\(m=(m+1)-1\); the marker-decoding lemma gives the required NO-instance
soundness.

Before the near-linear reduction, first use the existing fresh-singleton
padding to raise the source character count to \(N_0\). For the constant-factor
reduction, pad only if needed to meet its \(N\ge2\) condition. In either case,
adding \(p\) fresh singleton words changes the source set size from \(m\) to
\(m+p\) and its simplifiability threshold from \(m-1\) to \(m+p-1\),
preserving the answer. Apply the uniform-length padding afterward. Its new
source length is \((m+p+1)F\ge N_0\) in the near-linear case (and at least 2
in the constant-factor case), so the relevant parameter bounds apply using
this resulting length as \(N\). Both transformations have polynomial size.
This proves the restricted corollary.

\section{Compression Savings and Fixed-Alphabet Approximation Schemes}\label{sec:utily}
The objectives of minimizing token count and maximizing savings have the same exact optimal vocabularies, yet different multiplicative approximation properties. In particular, if $U_D(V)\ge(1-\varepsilon)\Uopt$, then $T_D(V)\le\Topt+\varepsilon(M-\Topt)$, which need not be a constant-factor approximation to $\Topt$.

\subsection{Compression-savings objective}
Inputs $s_i$ are explicit independent strings over an alphabet of size $\sigma$, with positive integer weights $w_i$ encoded in binary. Every character is mandatory. A vocabulary $V$ contains at most $K$ optional literal types, and only actual substrings of length at least two are useful. Write $n=\sum_i|s_i|$, $M=\sum_iw_i|s_i|$, and $I$ for full encoding length. DIRECT chooses a minimum-token parse; longest match (LM) repeatedly emits the longest matching available prefix. Let $T(V)$ be weighted token count and $U(V)=M-T(V)$ its savings utility. In this section we maximize $U$: a multiplicative savings guarantee is not a multiplicative token-count guarantee. Positions in different inputs remain separate, including equal inputs.

\subsection{Occurrence coverage}
The maximum-coverage relaxation and its greedy implementation were studied
by \citet{lim2025}. We state the relation between coverage and the savings
of a valid DIRECT parse explicitly; this is the step needed to convert a
coverage guarantee into a guarantee for the original objective.
An occurrence $[a,b)$ saves $b-a-1$ tokens and covers integer boundaries $a+1,\ldots,b-1$. Give each boundary weight $w_i$. Let $C(V)$ be the union weight of boundaries covered by all actual occurrences of selected types. It is a monotone submodular coverage function.
\begin{lemma}For DIRECT, $C(V)/2\le U(V)\le C(V)$, and the factor two is sharp.\end{lemma}
\begin{proof}
The boundaries saved by a parse belong to the covered union, proving the upper bound. On one input take an inclusion-minimal subfamily covering the same boundary union. Sort its intervals $[a_j,b_j)$ by start. Starts and ends are strictly increasing, since containment would make an interval redundant. If $a_{j+2}<b_j$, the integer boundary sets of intervals $j$ and $j+2$ overlap or are consecutive. Their union then covers every boundary of interval $j+1$, a contradiction. Thus $b_j\le a_{j+2}$. Odd intervals form a character-disjoint family, as do even intervals. Their total saving is at least the union size, so one family saves at least half. Fill gaps with characters and apply this per input, preserving weights. For $s=0^{2l-1},V=\{0^l\}$, coverage is $2l-2$ and utility $l-1$.
\end{proof}
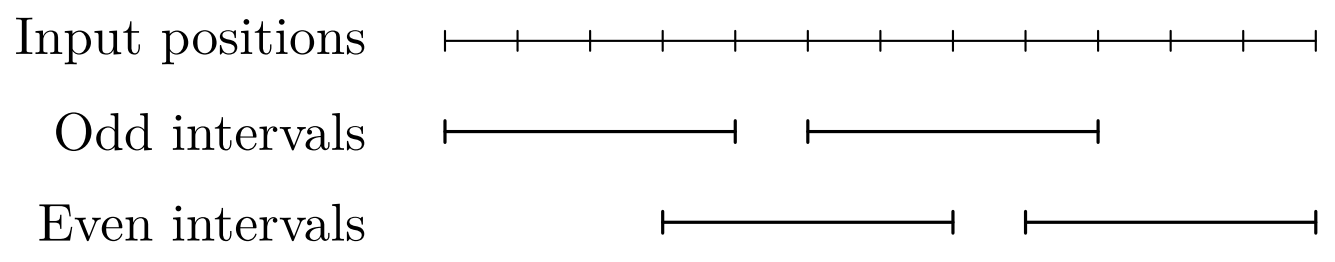
\begin{figure}[H]
\centering
\begin{tikzpicture}[x=0.48cm,y=0.60cm,line cap=round]
  \node[anchor=east,font=\small] at (-0.8,2.0) {Input positions};
  \node[anchor=east,font=\small] at (-0.8,1.0) {Odd intervals};
  \node[anchor=east,font=\small] at (-0.8,0.0) {Even intervals};
  \draw[thin] (0,2)--(12,2);
  \foreach \x in {0,...,12} {\draw[thin] (\x,1.89)--(\x,2.11);}
  \foreach \a/\b in {0/4,5/9} {
    \draw[semithick] (\a,1)--(\b,1);
    \draw[semithick] (\a,0.88)--(\a,1.12);
    \draw[semithick] (\b,0.88)--(\b,1.12);
  }
  \foreach \a/\b in {3/7,8/12} {
    \draw[semithick] (\a,0)--(\b,0);
    \draw[semithick] (\a,-0.12)--(\a,0.12);
    \draw[semithick] (\b,-0.12)--(\b,0.12);
  }
\end{tikzpicture}
\caption{Schematic illustration of two groups of nonoverlapping intervals from a minimal occurrence cover. Each group yields a feasible parse after unused positions are filled with character tokens.}
\label{fig:interval-groups}
\end{figure}
\begin{theorem}Coverage greedy achieves $U(V)\ge(1-e^{-1})\OPT_U/2$ in deterministic polynomial time, over arbitrary alphabets and even a restricted supplied candidate pool.\end{theorem}
\begin{proof}
For optimal $K$-type coverage $C^*$, the residual at step $j$ is at most the sum of the marginals of its at most $K$ types. A maximum marginal therefore removes at least $(C^*-C_j)/K$. Iteration gives $C_K\ge[1-(1-1/K)^K]C^*$. As $C^*\ge\OPT_U$, the lemma proves the claim. Zero residual permits early stopping. There are polynomially many actual substring types and boundaries; bitset coverage and exact integer-weight comparisons are polynomial. Evaluate the returned dictionary by position dynamic programming. For $K=0$ utility is zero; $K=1$ may instead be optimized exactly.
\end{proof}
\subsection{True-marginal greedy}
Unlike the occurrence-based scoring used by GreedTok~\citep{lim2025,baliga2026},
the algorithm analysed here evaluates every candidate using its true
increment in optimal DIRECT savings, recomputing the shortest parses after
adding each candidate type.
\begin{lemma}For any two DIRECT dictionaries $A,O$,
 $\displaystyle U(O)\le3U(A)+\sum_{t\in O\setminus A}\bigl(U(A\cup\{t\})-U(A)\bigr).$ 
\end{lemma}
\begin{proof}
Fix optimal parses $P,Q$ for $A,O$. Discard from $Q$ optional intervals contained in one optional $P$ interval. Their total saving is at most $U(A)$, because within a $P$ interval $I$ the disjoint discarded intervals save at most $|I|-1$. For each type $t$ insert all its surviving $Q$ intervals into $P$, removing intersecting optional $P$ intervals and filling gaps with characters. This feasible parse proves that the marginal of $t$ is at least its surviving saving minus the removed saving. This remains true for $t\in A$, whose marginal is zero by optimality of $P$. Each $P$ interval meets at most two surviving $Q$ intervals: one may cross its left endpoint and one its right, and any other intersecting interval would be contained. Summing losses over types charges $P$ at most twice. Adding the discarded saving proves the inequality.
\end{proof}
\begin{theorem}Greedy using the true DIRECT marginal utility achieves $(1-e^{-3})/3$ of optimum, uniformly over $K\ge1$.\end{theorem}
\begin{proof}
At a dictionary $A$, some available marginal is at least $(\OPT_U-3U(A))/K$ when this is positive; monotonicity gives the same inequality otherwise. For $K\ge3$, iteration from zero yields $U(G_K)\ge[1-(1-3/K)^K]\OPT_U/3$. For $K=2$, partition the optimal parse saving by its two types; the best singleton realizes at least half. For $K=1$ it is exact. Exhausted candidates give the optimum. Every true marginal is computed by an exact segmentation DP, so the algorithm is polynomial.
\end{proof}
The coefficient three in the exchange inequality cannot be reduced universally. On a string of $3L+1$ distinct characters allow intervals $A=[L,2L+1)$, $B=[0,L+1)$, $C=[L+1,2L)$, and $D=[2L,3L+1)$. Dictionary $\{A\}$ saves $L$; each of the three other singleton additions has zero marginal, while $\{B,C,D\}$ saves $3L-2$. This concerns a restricted pool and the inequality, not a tight unrestricted-greedy ratio.
\subsection{Additive support structure}
For any full segmentation $P$, let $a_P(t)$ be its weighted number of $t$ occurrences times $|t|-1$. Expanding every token outside $A$ into characters proves
 $\displaystyle U(A)=\max_P\sum_{t\in A}a_P(t).$ 
Thus utility is XOS (a maximum of nonnegative additive functions), monotone and subadditive. An optimal parse supplies a tight lower support. Submodularity does not follow; the exchange proof above is the stronger interval-specific information needed here.

\subsection{Fixed-alphabet approximation schemes}\label{sec:eptas}
We strengthen the savings guarantees using short literals and a deterministic,
approximately lossless candidate pool. The result concerns savings, not the
direct token-count ratio. In contrast to the unrestricted-alphabet
inapproximability in \citet{baliga2026}, our approximation scheme assumes
that the alphabet size is fixed.

\subsection{A pointwise PTAS for DIRECT and longest match}
Fix rational $0<\epsilon<1$, set $b=\lceil1/\epsilon\rceil$ and $J=\sum_{j=2}^b\sigma^j$. If $K<J$, enumerate all subsets of at most $K$ actual substring types and evaluate exactly under the relevant parser. For fixed $\sigma,\epsilon$ this is polynomial. If $K\ge J$, take every occurring type of lengths $2$ through $b$. Both parsers then use at most $\lceil l/b\rceil$ tokens on a length-$l$ input, and LM uses exactly that many. Thus
 $\displaystyle l-\lceil l/b\rceil=(l-1)-\lfloor(l-1)/b\rfloor\ge(1-\epsilon)(l-1).$ 
Every feasible per-input saving is at most $l-1$, so positive weighted summation proves a PTAS. More generally the pointwise inequality proves a PTAS for any finite nonnegative, coordinatewise monotone, degree-one homogeneous aggregation $G$ of the per-input savings vector, provided attainable objective values have an exact polynomial comparator. This extension does not assert the EPTAS below for such $G$.
\subsection{A small deterministic pool for DIRECT}
Discard empty and singleton inputs; let $M_0$ be remaining weighted length. For $K\ge1$, some bigram has boundary coverage at least $(M_0-W_0)/\sigma^2$, where $W_0\le M_0/2$ is weighted input count. Alternating its occurrence edges on each path gives disjoint bigrams saving half the coverage. Hence
 $\displaystyle U^*\ge M_0/(4\sigma^2).$ 
For each literal type $t$, let $R_t$ be the union of its occurrence character positions. Retain ranges of weight at least $\delta M_0$, with $\delta=\epsilon/(8K\sigma^2)$. There are at most $n(n+1)/2$ ranges. Greedily select the position incident to most unhit retained ranges. If $r$ remain, weighted averaging gives a position incident to at least $\delta r$: indeed $\sum_p w(p)c(p)\ge r\delta M_0$. Thus $O(\delta^{-1}\log(n+2))$ positions hit all retained ranges, deterministically.

Put $\eta=\epsilon/16$. At every chosen position $p$ and every dyadic scale $h\le n$, use endpoints on $p+s\mathbb Z$ within distance $2h$ and within its input, where $s=\max(1,\lfloor\eta h\rfloor)$. Endpoints include valid terminal coordinates. Include every resulting substring of length at least two. There are $O(\eta^{-2})$ intervals per scale: if $s=1$ then $h<2/\eta$, and otherwise $s\ge\eta h/2$. The pool has size
 $\displaystyle O(K\sigma^2\epsilon^{-3}\log^2(n+2)).$ 
\begin{lemma}This pool contains a dictionary of at most $K$ types with DIRECT utility at least $(1-\epsilon)U^*$.\end{lemma}
\begin{proof}
Fix optimal parses. Removing all types with light ranges loses at most $K\delta M_0\le\epsilon U^*/2$. Every other type has an occurrence containing a hit position $p$. For its length $l$ choose $h\le l<2h$, and round both occurrence endpoints inward on the grid. If $s=1$ nothing changes. Otherwise total loss is less than $2s\le2\eta l\le4\eta(l-1)$, so the retained interval has length at least two and saving at least $(1-4\eta)(l-1)$. Record its fixed literal prefix and suffix offsets and use the same trimmed word inside every old occurrence. Equal types have equal retained literals. Since original token intervals were disjoint, the replacements plus characters are legal. Each surviving old type produces at most one new type, so budget is preserved; collisions only help. Savings are at least $(1-\epsilon/4)(1-\epsilon/2)U^*\ge(1-\epsilon)U^*$.
\end{proof}
Enumerate at-most-$K$ subsets of the pool and evaluate by exact parsing. Writing its size as $A\log^2(n+2)$ with parameter-only $A$, use $x^{2K}e^{-x}\le(2K/e)^{2K}$ at $x=\log(n+2)$ to absorb the logarithmic power into $f(K,\sigma,1/\epsilon)(n+2)$. All substring, range-incidence, and exact-weight operations have absolute polynomial overhead in the full encoding, including rational accuracy bits. Loops can be capped at the actual positions and candidates.
\begin{theorem}DIRECT utility has a deterministic approximation scheme in time $f(K,\sigma,1/\epsilon)I^{O(1)}$, and an EPTAS $g(\sigma,1/\epsilon)I^{O(1)}$ for fixed alphabet.\end{theorem}
\begin{proof}
The first assertion is the pool enumeration. For the second, use the universal dictionary when $K\ge J$; otherwise $K<J$ is bounded by $\sigma,\epsilon$, and maximize the parameter factor over this finite range.
\end{proof}
The replacement argument proves no LM EPTAS: LM may ignore the displayed replacement parse. Neither scheme gives exact FPT in $K$. There is no deterministic FPTAS for either binary utility model unless $\mathrm P=\mathrm{NP}$: on explicit unit-weight inputs take $\epsilon=1/(M+1)$. Integral utility bounded by $M$ must then be optimal, contradicting exact hardness. Encoded large weights are not needed for this exclusion.

\section{Subexponential Exact Algorithms}\label{sec:exact}
The next results exploit representative occurrences of chosen token types to reduce the search space for exact optimization. The bounds are measured in the total length of distinct input strings.

\subsection{Model and encoding}
An instance is an explicit list of binary strings, a nonnegative integer budget $K$, and mandatory character vocabulary $\{0,1\}$. All asymptotic logarithms are base two; integral estimates use natural logarithms. At most $K$ additional word types may be selected. Each listed string is parsed optimally, and the objective is the sum of token counts, including repeated entries with multiplicity. Empty strings have cost zero and are discarded.

Let $q$ be the number of distinct nonempty input strings and
 $\displaystyle n=\sum_{i=1}^q |s_i|$ 
their total length counting each distinct string once. Let $I$ be the full input representation length, including strings, multiplicities or weights, and budget metadata. Search is measured in $n$; input reading and arithmetic are polynomial in $I$. For an ordinary explicit corpus of expanded character count $M$ and polynomial metadata, $I=M^{O(1)}$.

\begin{theorem}[Unconditional binary DIRECT]
There is an exact deterministic algorithm with running time
 $\displaystyle 2^{O(n\log\log n/\log n)}I^{O(1)}=2^{o(n)}I^{O(1)}.$ 
For $K\ge1$, the same algorithm also has a $K$-sensitive endpoint-enumeration bound $2^{O(K\log(1+n/K))}I^{O(1)}$. The exponent constants and polynomial exponent are absolute. For an ordinary explicitly expanded corpus the total time is $2^{o(M)}$. Unary instances have the sharper exact bound $2^{O(\sqrt n)}I^{O(1)}$.
\end{theorem}

\subsection{Canonical parses give disjoint representatives}
Group identical nonempty inputs into distinct strings $s_1,\ldots,s_q$ with positive multiplicities $a_i$. For any dictionary $V$, the corpus cost is $\sum_i a_i\operatorname{tok}_V(s_i)$, where $\operatorname{tok}_V(s_i)$ is the minimum number of vocabulary tokens in a parse. One canonical optimum parse per distinct input suffices.

The empty corpus is solved with no optional words, and $K=0$ uses characters. If $K\ge q$, selecting every distinct input of length at least two as a word makes each nonempty copy cost one token, which is optimal. The remaining case has $1\le K<q$.

\begin{lemma}[Disjoint representatives]\label{exact:lem:rep}
There is an optimal optional dictionary $V$ with one representative occurrence for each type, such that these intervals are pairwise disjoint and $|V|\le K$, $\sum_{w\in V}|w|\le n$. Every optional type may be taken to have length at least two and to occur as an input substring.
\end{lemma}
\begin{proof}
Choose one minimum-token parse of each distinct input under an optimal vocabulary and delete optional types unused by these fixed parses. This preserves the chosen parses and their weighted objective. For each remaining type choose one occurrence from its fixed parses. Token intervals within a parse are disjoint, and distinct inputs occupy disjoint portions of the flattened corpus, so the selected intervals are pairwise disjoint and have total length at most $n$. Each is a used substring; optional characters are redundant.\end{proof}

This chooses occurrences from one fixed parse collection. Arbitrary substring occurrences may overlap; the chosen token occurrences cannot.

\subsection{Endpoint enumeration and exactness}
Flatten the distinct inputs into $n$ character positions while retaining boundaries. For $k=0,\ldots,\min(K,\lfloor n/2\rfloor)$, enumerate every increasing $2k$-position set
 $\displaystyle p_1<q_1<p_2<q_2<\cdots<p_k<q_k.$ 
Pair consecutive endpoints into inclusive intervals $[p_j,q_j]$. Reject intervals crossing an input boundary. Read the interval contents, deduplicate equal words, and evaluate their set as an optional dictionary of size at most $K$. Adjacent intervals are allowed when $q_j+1=p_{j+1}$; the inclusive intervals then share no character.

Evaluate each candidate by exact segmentation dynamic programming on each distinct input and weight by its multiplicity. Every candidate is feasible. Conversely, Lemma~\ref{exact:lem:rep} supplies an optimal dictionary with disjoint inclusive representatives; sorting their endpoints yields an enumerated selection with exactly those interval contents. Thus the minimum evaluated cost is optimal.

The number of endpoint selections is
 $\displaystyle S(n,K)=\sum_{k=0}^{\min(K,\lfloor n/2\rfloor)}\binom{n}{2k}.$ 
For $K<n/4$, the binomial lower-tail bound through $2K$ gives $S(n,K)\le(en/(2K))^{2K}$. For $K\ge n/4$, use $S(n,K)\le2^n$ and monotonicity of $x\log(1+n/x)$ to obtain the same asymptotic conclusion:
 $\displaystyle S(n,K)\le2^{O(K\log(1+n/K))}.$ 
The constants are absolute. Endpoint generation, boundary tests, deterministic content deduplication, and exact segmentation take $I^{O(1)}$ time per branch with an absolute polynomial exponent. Weighted arithmetic is polynomial in $I$. A depth-first enumeration avoids any exponential-space requirement. Check $K$ against $q$ first, so a large binary-encoded budget causes no loop proportional to its numeric value.

\subsection{Binary counting gives a subexponential bound}
The nontrivial branch has $K<q$, where $q$ counts distinct strings. For $q\ge16$, fewer than $q/2$ nonempty binary strings have length below $\lfloor\log_2q\rfloor-1$. Therefore
 $\displaystyle n\ge(q/2)(\log_2q-2)\ge(q/4)\log_2q.$ 
For $n\ge256$, if $q\le\sqrt n$ then $q\le n/\log_2n$; otherwise the preceding inequality gives $q\le8n/\log_2n$. Hence $q=O(n/\log n)$. Since $x\log(1+n/x)$ is increasing,
 $\displaystyle K\log(1+n/K)\le q\log(1+n/q)=O(n\log\log n/\log n).$ 
Substitution into the endpoint bound proves the endpoint theorem. For ordinary expanded length $M$, $n\le M$ and $I=M^{O(1)}$, so the complete running time is $2^{o(M)}$. For encoded multiplicities or unusual metadata, the $I^{O(1)}$ factor is retained; the theorem does not claim subexponential input-reading in $n$ alone.

\subsection{Hall matching and a second exact search}
Join a used optional type to each distinct input whose fixed canonical parse uses it.
\begin{lemma}[Distinct-input representatives]\label{exact:lem:hall}
This bipartite graph has a matching saturating all used optional types.
\end{lemma}
\begin{proof}
If a type set $A$ has fewer neighboring inputs $N(A)$ than types, delete $A$ and add each whole input in $N(A)$ as a word when it is not already mandatory or retained. The budget decreases because $|N(A)|<|A|$. Inputs outside $N(A)$ retain their parses. Each input in $N(A)$ now costs one token. All types of $A$ were used in these inputs, so their parses contain at least $|A|$ token occurrences across fewer than $|A|$ inputs. At least one affected input had more than one token; the positive multiplicities therefore make the change strictly improve the objective, a contradiction. Hall's theorem gives the matching.\end{proof}

Thus one may independently choose, for each distinct input, either no word or one length-at-least-two substring, then deduplicate, enforce the $K$ cap, and evaluate. If $\ell_i=|s_i|$, the number of candidates is at most
 $\displaystyle \prod_i(1+\ell_i^2)\le(1+(n/q)^2)^q=2^{O(q\log(1+n/q))},$ 
by Jensen's inequality for the concave function $\log(1+x^2)$ on $x\ge1$. This offers a second exact search; the endpoint search remains the stated universal bound.

\subsection{Unary denomination sets}
When all strings are powers of one character, optional words are determined by lengths $\ell\ge2$. Lemma~\ref{exact:lem:rep} gives an optimal distinct length set $A$ with $|A|\le K$ and $\sum_{\ell\in A}\ell\le n$. Enumerate these increasing sets and evaluate each input length by the minimum-coin DP using denominations $A\cup\{1\}$.

Let $C_n$ count all sets of integers at least two with sum at most $n$. For $x=e^{-1/\sqrt n}$,
 $\displaystyle C_nx^n\le\prod_{j\ge2}(1+x^j),\qquad \log C_n\le\sqrt n+\sum_{j\ge2}e^{-j/\sqrt n}\le2\sqrt n.$ 
An increasing-set recursion visits each feasible set once, and every prefix is feasible; work per node is polynomial. The infinite product is only an analysis bound. This proves the unary $2^{O(\sqrt n)}I^{O(1)}$ result. Inputs are explicit; binary-encoded unary lengths are not treated as free expansions.

\subsection{Structural parameters for exact algorithms}\label{sec:structure}
\subsection{Cross-input sharing and private whole words}
Let $\ell$ be the longest substring shared by two \emph{different distinct} nonempty inputs, with $\ell=0$ when none exists. Group duplicate inputs into positive weights before computing $\ell$; repetition within one input does not count. Pairwise longest-common-substring dynamic programming computes $\ell$ in $I^{O(1)}$ time.

Private types occurring in only one distinct input can be replaced by that whole input: remove all such selected types, add each affected whole input once, and parse it in one token. The number of additions is at most the number of removed types. This preserves an optimum. For a fixed shared dictionary, private whole inputs yield independent weighted gains.\par
\subsection{Shared-type scope width}
Call a length-at-least-two optional type \emph{shared} when it occurs in at least two distinct inputs, and let $U$ be the number of actual shared types. Every other useful type is private. Replacing every used private type by its whole input, as in the private-word normalization, shows that some optimum consists of shared types and private whole inputs. Enumerate the $2^U$ shared subsets $S$ and, for each, select the best $K-|S|$ private whole-input gains after exact weighted segmentation. This gives another exact time bound $2^U I^{O(1)}$.

Fix an order of distinct inputs and flatten their positions without permitting a token to cross an input boundary. For each \emph{shared} type $w$, let $f_w$ be its first occurrence start and $l_w$ its last occurrence end. At position $x$ put
 $\displaystyle \operatorname{Live}(x)=\{w\text{ shared}:f_w\le x\le l_w\}, \qquad p=\max_x|\operatorname{Live}(x)|\le U.$ 

\begin{theorem}[Shared-type scope-width exact algorithm]
For this supplied ordering, exact weighted DIRECT is solvable in $(K+1)2^p I^{O(1)}$ time. After the polynomial $K\ge q$ shortcut, $K<q\le n$, so this is FPT in $p$ with absolute polynomial overhead. No bound in $K$ alone follows.
\end{theorem}
\begin{proof}
At each token boundary $x$, keep a state $(x,S,k)$: the minimum cost of a partial corpus parse, its number $k$ of distinct optional types used so far, and exactly those used \emph{shared} types still in $S\subseteq\operatorname{Live}(x)$. A character edge adds its input weight. A shared-type occurrence edge adds that weight and increments $k$ only when its type is absent from $S$. A private whole-input edge adds its input weight, increments $k$ once, and needs no mask bit because that whole word has only one occurrence in the distinct corpus; length-one inputs instead use their free character token. Project the updated shared mask onto $\operatorname{Live}(y)$ at the edge's endpoint $y$. Input-boundary edges cost zero. A previously used shared type no longer live cannot appear again, while a type first becoming live during a jumped-over interval has not been used. Thus the state simulates all feasible parse sequences and charges each type once. There are at most $(n+1)(K+1)2^p$ states and polynomially many occurrence transitions per state. Therefore the dynamic program has the stated running time.
\end{proof}

For the \emph{decision} version with a weighted total-token threshold, actual optional substrings have polynomially many candidates. A polynomial-size Boolean circuit unrolls exact segmentation and threshold comparison from activation variables. Padding an at-most-$K$ chosen set with $K$ ignored variables reduces the nontrivial $K<q$ branch to weight-exactly-$K$ circuit satisfiability, so decision DIRECT belongs to W[P]. This is an upper bound in parameterized complexity; it does not establish FPT or W-hardness for $K$ alone.

\section{Unary Bottom-Up and Optimal Pair Encoding}\label{sec:unaryapx}
We first establish APX-hardness of unary standard bottom-up tokenisation, then prove strong NP-completeness for unary optimal pair encoding (OPE). The difference is whether merge rules are executed on current parses or used only to build a vocabulary for a subsequent optimal parse. The definitions of both models were given in Section~\ref{sec:preliminaries}.

\subsection{Hardness theorem and occurrence alignment}
For unary input $a^N$, we identify each token with its length. A rule merges all nonoverlapping occurrences of one adjacent pair, from left to right. We use at most $K$ rules, and the cost is the number of tokens in the parses they produce.

\begin{theorem}[L-reduction from cubic Vertex Cover]\label{thm:apx-L}
Unary standard bottom-up token-count minimisation is APX-hard. More precisely, from a simple cubic graph $G$ we can build an explicitly represented unary dataset $D$, a budget $K$, and a baseline $B_0$, in polynomial time, such that
 $\displaystyle \OPT_{\BU}(D,K)=B_0+m+\tau(G),$ 
where $m=|E(G)|$ and $\tau(G)$ is the minimum vertex-cover size. Every feasible bottom-up sequence with token cost $L$ can be decoded to a vertex cover $C_G$ satisfying
 $\displaystyle |C_G|-\tau(G)\le L-\OPT_{\BU}(D,K),$ 
and $\OPT_{\BU}(D,K)\le100000\,\tau(G)$.
\end{theorem}

Write $\nu_2(z)$ for the $2$-adic valuation of a positive integer $z$. We use the following invariant: if $p=2^s$ is the largest power of two productively created by doubling, each current unary parse has form
 $\displaystyle p,p,\ldots,p,t_1,\ldots,t_r,\qquad s>\nu_2(t_1)>\cdots>\nu_2(t_r)\ge0.$ 
The prefix may be empty. For a non-doubling rule the operand of smaller valuation appears uniquely per parse, so the rule can act at most once on each individual string. A doubling replaces adjacent equal $p$ tokens in the prefix by $2p$ and preserves the prescribed tail form.

\begin{lemma}[Occurrence alignment]\label{lem:align}
After at most $K$ productive bottom-up merge rules, every actual occurrence of a token of length $w$ begins at a position divisible by
 $\displaystyle d_K(w)=\max\left\{1,2^{\left\lfloor\log_2(w/(K+1))\right\rfloor}\right\}.$ 
In particular, if a string of length $N$ ends with a two-token parse $(x,y)$, then $x\equiv0\pmod{d_K(y)}$ and $N\equiv y\pmod{d_K(y)}$.
\end{lemma}
\begin{proof}
Contract maximal subtrees consisting only of power-of-two doublings in the merge tree of an actual token occurrence. Let $\Lambda$ be the largest contracted leaf length. By induction over the merge sequence, the largest pure-doubling leaf labels of consecutive current tokens are nonincreasing from left to right, and each actual occurrence begins at a multiple of its own largest such label. A non-doubling rule combines a left label at least as large as its right label and retains the left occurrence start; a doubling operates on the pure-power prefix and creates $2p$ tokens aligned to multiples of $2p$. An actual token whose total length is a power of two cannot be created by merging operands of distinct $2$-adic valuations, so the claimed dichotomy is exhaustive.

Each contracted internal merge-tree node is a non-doubling event. By the parse invariant, no such rule acts twice on the same string; hence a token formed after at most $K$ rules has at most $K$ non-doubling events and at most $K+1$ leaves. Its length is thus $w\le(K+1)\Lambda$. Both $\Lambda$ and $d_K(w)$ are powers of two, so $d_K(w)$ divides $\Lambda$, implying the desired alignment. The congruences for a final two-token parse follow by applying alignment to its right token.
\end{proof}

\subsection{Cubic vertex cover and twelve separated bands}
Let $G$ be simple cubic with $n\ge4$ vertices, $m=3n/2$ edges and minimum cover size $\tau(G)$. The source problem is APX-hard \citep{alimontikann2000}. Subdivide every original edge $uv$ into the path $u-b_e-c_e-v$ to form $H$. Then
 $\displaystyle N=|V(H)|=n+2m=4n,\quad m_H=|E(H)|=3m,\quad \tau(H)=m+\tau(G).$ 
Indeed, a cover of $G$ extends by choosing one appropriate internal vertex per original edge. Conversely, if $C_H$ covers $H$ and $S=C_H\cap V(G)$, every edge of $G$ not covered by $S$ requires both internal vertices. Adding one endpoint of each uncovered original edge gives a cover $C_G$ with $|C_G|\le |C_H|-m$.

Properly colour $G$ with colours $1,\ldots,4$ and orient every edge from its lower-coloured endpoint $u$ (colour $c$) to its higher-coloured endpoint $v$ (colour $d>c$). Use twelve exponent bands
 $\displaystyle b(X_c)=14-3c,\qquad b(Y_c)=13-3c,\qquad b(Z_c)=12-3c.$ 
Set $M_0=N+2$ and choose
 $\displaystyle q=2^{\lceil\log_2(10000M_0^2)\rceil},\qquad \ell=\log_2q.$ 
Give distinct vertex IDs in $\{1,\ldots,n\}$ and edge IDs in $\{1,\ldots,m\}$. Define, for an original vertex $v$ of colour $c$ and an edge $e=(u,v)$ oriented from colour $c$ to $d>c$,
\[
\begin{aligned}
x_v&=(M_0+\operatorname{id}(v))q^{b(X_c)},\\
y_v&=(M_0+\operatorname{id}(v))q^{b(Y_c)},\\
z_e&=(M_0+\operatorname{id}(e))q^{b(Z_c)}.
\end{aligned}
\]
All atom coefficients lie in $[M_0+1,2M_0]$. Each atom occurs in at most six of the ordered triples
 $\displaystyle (x_u,y_u,z_e),\quad(y_u,z_e,x_v),\quad(z_e,x_v,y_v).$ 
These triples encode respectively the edges $ub_e$, $b_ec_e$ and $c_ev$ of $H$. The real choice length associated with an original vertex $v$ is $R_v=x_v+y_v$; the two internal vertices have $R_{b_e}=y_u+z_e$ and $R_{c_e}=z_e+x_v$.

Their band pairs are precisely $(X_c,Y_c)$, $(Y_c,Z_c)$ and $(Z_c,X_d)$ for $c<d$. In contrast, the nonadjacent band pairs within triples are $(X_c,Z_c)$, $(Y_c,X_d)$ and $(Z_c,Y_d)$. Thus no real choice coincides with the sum of nonadjacent atoms of a triple.

\subsection{Base lengths, explicit input, and canonical merges}
Put
 $\displaystyle F=\{s q^b:1\le s\le2M_0,\ 0\le b\le13\} \cup\{2^r:0\le r\le14\ell\},\qquad h=|F|-1.$ 
The character length $1$ is free. There are $N$ dummy choice lengths $D_i=q^{13}+(M_0+i)q^{12}$, one for each $1\le i\le N$. Let
 $\displaystyle C=2^{18},\quad A=2^{20},\quad K=h+N.$ 
Construct the explicit dataset using $A$ copies of every non-character base anchor $a^f$ for $f\in F\setminus\{1\}$, $C$ copies of each real-choice anchor $a^{R_w}$, $C+1$ copies of every dummy anchor $a^{D_i}$, and one copy of each of the $m_H$ triple-sum edge strings. All string lengths and repetitions are written in unary. Distinct categories have different digit supports in radix $q$.

The construction gives $h\le67M_0$, hence $K+1\le68M_0-1$, while
 $\displaystyle q>16M_0(K+1),\qquad q<20000M_0^2.$ 
Every input length is polynomial in $n$: the maximum anchor length is $q^{14}=O(n^{28})$, and there are $O(n)$ strings, giving explicit input size $O(n^{29})$.

A common preamble of exactly $h$ rules constructs every member of $F\setminus\{1\}$. First double powers of two up to $q^{14}$. Every parse is then the descending binary expansion of its length. For each band $b$ in increasing order and each non-power-of-two coefficient $s\le2M_0$, merge
 $\displaystyle (2^{\lfloor\log_2s\rfloor}q^b,\ (s-2^{\lfloor\log_2s\rfloor})q^b) \mapsto sq^b$ 
when the second operand is nonzero. Smaller coefficients have been assembled first, so the rules assemble the corresponding suffixes and then the leading power-of-two. As $q>2M_0$, the rules do not cross band boundaries. After this preamble every base anchor is a singleton; each real and dummy anchor has its intended two-token parse; and every edge string has the corresponding three-atom parse.

For $S\subseteq V(H)$ of cardinality $t$, append the $t$ real choice rules for vertices of $S$ and $N-t$ different dummy rules. The resulting cost is
 $\displaystyle B_0+t+u_H(S),\qquad B_0=Ah+(3C+1)N+2m_H,$ 
where $u_H(S)$ counts edges of $H$ not covered by $S$. An incident choice rule reduces an edge triple from three tokens to two; overlapping incident rules cannot both act within the same triple.

\subsection{Unrestricted helpers and the global lower bound}
Call a token length \emph{normal} if it belongs to $F$, is a real choice, or is a dummy choice. All other non-character lengths are helpers. The radix construction implies that the only representations of an edge-string length as a sum of two normal lengths are its two intended incident-choice-plus-atom representations. Briefly, two single-band normal lengths cannot represent a three-digit target; a real choice plus a base can do so only by using the two intended adjacent bands; two real choices either overlap with excessive coefficient or occupy too many bands; and a dummy is too large.

The difficult step is to control two-token parses created by \emph{arbitrary} additional helper lengths. Consider an edge string with no generated incident real choice, of length
 $\displaystyle T=a q^i+bq^j+cq^k=A_0+B'_0+C_0,\qquad i>j>k, \quad M_0+1\le a,b,c\le2M_0.$ 
Suppose its actual final parse consists of $(x,y)$. If $y<q^{k+1}$, then $a q^i<x<(a+1)q^i$. Were $x$ normal, it would have to equal the left intended incident choice $A_0+B'_0$, contradicting the assumption. Charge this edge to the helper $x$; its leading digit pins its top atom.

If $y\ge q^{k+1}$, let $P_k$ be the least power of two greater than $2M_0q^k$. Then
 $\displaystyle 2M_0q^k<P_k\le4M_0q^k,\qquad P_k\mid q^{k+1},\qquad P_k\le d_K(y).$ 
Occurrence alignment gives $y\equiv T\equiv cq^k\pmod{P_k}$. If $y$ is a helper, charge the edge to $y$; for fixed bottom band $k$, its residue pins the bottom atom. Otherwise a normal $y$ cannot be a one-band base or a dummy. It must be a real choice whose lower atom equals $C_0$ and whose upper band is strictly between $k$ and $i$. Its complement $x$ is then a helper satisfying
 $\displaystyle |x-aq^i|\le2M_0q^{i-1}<\tfrac12q^i.$ 
Charge the edge to $x$; its leading digit pins one of at most two possible top atoms per band.

For each fixed helper, there are at most two possible top atoms per band and one possible bottom atom per band. With twelve bands and at most six edge-string appearances per atom, a helper is charged for at most $(2\cdot12+12)\cdot6=216$ relevant two-token edge strings. A singleton edge string requires its own distinct edge-length helper. Allowing for two saved tokens on such singleton strings yields the conservative saving bound of $218$ per distinct helper.

Consequently, if $S$ is the set of generated real-choice lengths and $r$ is the number of distinct helper lengths, the edge-string token count obeys
 $\displaystyle L_{\mathrm{edges}}\ge3m_H-\operatorname{cov}_H(S)-218r =2m_H+u_H(S)-218r.$

\subsection{Optimal value and L-reduction constants}
Suppose an arbitrary feasible sequence generates $t$ real-choice lengths, $d$ dummy lengths and $r$ helper lengths, while omitting $b$ of the $h$ non-character base lengths. Since at most $K$ distinct non-character lengths are generated,
 $\displaystyle t+d+r\le N+b.$ 
Each omitted base anchor costs at least one extra token per copy, and each omitted real or dummy choice anchor costs at least one extra token per copy. Combining the anchor bounds with the helper bound gives
\[
\begin{aligned}
L&\ge Ah+Ab+C(2N-t)+(C+1)(2N-d)+2m_H+u_H(S)-218r\\
&=B_0+t+u_H(S)+(C+1)(N-t-d)+Ab-218r\\
&\ge B_0+t+u_H(S)+(A-C-1)b+(C+1-218)r\\
&\ge B_0+t+u_H(S)+b+r.
\end{aligned}
\]
Add one endpoint for every edge of $H$ left uncovered by $S$, obtaining a vertex cover $C_H$ with
 $\displaystyle |C_H|\le t+u_H(S)\le L-B_0.$ 
Conversely, choosing the real choices of a minimum vertex cover of $H$ in the canonical construction attains $B_0+\tau(H)$, so
 $\displaystyle \OPT_{\BU}(D,K)=B_0+\tau(H)=B_0+m+\tau(G).$ 
Use the decoder for the subdivision step to obtain a cover $C_G$ of the original graph with
 $\displaystyle |C_G|-\tau(G)\le L-B_0-m-\tau(G) =L-\OPT_{\BU}(D,K),$ 
which is the L-reduction inequality with $\beta=1$. Finally, the construction bounds $h\le210n$, and cubic graphs satisfy $\tau(G)\ge n/2$. Substituting $A=2^{20}$, $C=2^{18}$, $N=4n$, and $m_H=9n/2$ yields
 $\displaystyle \OPT_{\BU}(D,K)\le 97662\,\tau(G)<100000\,\tau(G).$ 
This gives the L-reduction with $\alpha=100000$ and $\beta=1$. It follows that the problem has no PTAS unless $\mathrm P=\mathrm{NP}$. APX-hardness alone does not imply membership in APX.

\subsection{Strong NP-completeness of unary OPE}\label{sec:unaryope}
The next reduction studies the merge-extracted-vocabulary version of OPE. It is distinct from the standard bottom-up execution model and requires its own hardness argument.

\begin{theorem}\label{unaryope:thm:main}
The decision version of unary optimal pair encoding is strongly NP-complete on explicitly represented unary datasets. The reduction uses only polynomially bounded lengths, polynomially many explicitly repeated strings, and a linear merge budget in the source graph size.
\end{theorem}

\subsection{Source problem and construction}
We reduce from \emph{maximum $k$-vertex coverage on bipartite graphs}. Let
 $\displaystyle G=(U\mathbin{\dot\cup}V,E),\qquad U=\{u_1,\ldots,u_p\},\quad V=\{v_1,\ldots,v_r\},$ 
where $p,r\ge1$, $n=p+r$, and $m=|E|$. For a set $S\subseteq U\cup V$, let $\cov(S)$ be the number of edges with at least one endpoint in $S$, and define
 $\displaystyle \cov_k(G)=\max_{|S|=k}\cov(S),\qquad 0\le k\le n.$ 
The decision problem asks whether $\cov_k(G)\ge t$, with $0\le t\le m$. This problem is NP-complete: its NP-hardness is established by Joret and Vetta~\citep{joretvetta2015}, and a vertex set is a polynomial certificate. The ``at most $k$'' version is equivalent by padding the chosen set.

Let $B$ be the smallest power of two strictly larger than $r$, and put
 $\displaystyle B=2^{\lceil\log_2(r+1)\rceil},\qquad Q=4B.$ 
Define the fixed base lengths
\begin{equation}\label{unaryope:eq:base}
F=\{1,\ldots,r\}\cup\{B,2B\}\cup\{Qi:1\le i\le p\}.
\end{equation}
The three parts are disjoint, $|F|=n+2$, and
 $\displaystyle h=|F|-1=n+1$ 
is the number of non-character base words. Associate choice lengths with the vertices:
\begin{equation}\label{unaryope:eq:choices}
\ell_i=Qi+B\quad(u_i\in U),\qquad
\rho_j=B+j\quad(v_j\in V).
\end{equation}
For every edge $u_iv_j\in E$, define
\begin{equation}\label{unaryope:eq:edge}
e_{ij}=Qi+B+j.
\end{equation}
All base, choice, and edge lengths are distinct across these categories; edge lengths are distinct for distinct edges. This follows from quotient and remainder modulo $Q$, using $1\le j\le r<B$.

Set the repetition counts
\begin{equation}\label{unaryope:eq:weights}
C=2m+1,\qquad A=2nC+3m+1.
\end{equation}
The dataset consists of the following actual strings, not binary-encoded weights:
\begin{center}
\renewcommand{\arraystretch}{1.2}
\begin{tabular}{lll}
\toprule
String type & Length & Copies of each string\\
\midrule
Base anchor & $f\in F\setminus\{1\}$ & $A$\\
Vertex anchor & $\ell_i$ or $\rho_j$ & $C$\\
Edge string & $e_{ij}$ for $u_iv_j\in E$ & $1$\\
\bottomrule
\end{tabular}
\end{center}
Finally, choose
\begin{equation}\label{unaryope:eq:budget}
K=h+k,\qquad
\delta=Ah+C(2n-k)+3m-t.
\end{equation}
We will prove the exact identity
\begin{equation}\label{unaryope:eq:identity}
L^*_{\DIR}(D,K)=L^*_{\OPE}(D,K)=L^*_{\BU}(D,K)
=Ah+C(2n-k)+3m-\cov_k(G).
\end{equation}

\subsection{Excluding unintended two-token representations}
Let the full set of potentially intended words be
 $\displaystyle \mathcal V=F\cup\{\ell_i:1\le i\le p\}\cup\{\rho_j:1\le j\le r\}.$ 
This is not a restriction on the input vocabulary. The lower-bound argument below establishes the stated form for optimal solutions of the constructed instance.

\begin{lemma}\label{unaryope:lem:arithmetic}
For every $i,j$, the only unordered representations of $e_{ij}$ as a sum of two elements of $\mathcal V$ are
 $\displaystyle e_{ij}=\ell_i+j=Qi+\rho_j.$ 
Moreover, $e_{ij}\notin\mathcal V$.
\end{lemma}
\begin{proof}
The low elements of $\mathcal V$, those smaller than $Q$, are
 $\displaystyle 1,\ldots,r,\ B,\ 2B,\ B+1,\ldots,B+r,$ 
and all are at most $2B$. The high elements are $Qt$ and $Qt+B$, for $1\le t\le p$.

Two low elements sum to at most $4B=Q<e_{ij}$. Two high elements have sum congruent to $0$, $B$, or $2B$ modulo $Q$, whereas $e_{ij}$ has remainder $B+j$, strictly between $B$ and $2B$. Thus precisely one summand is high.

If the high summand is $Qt$, the other summand must be $B+j$, and $t=i$, since the low summand is smaller than $Q$. If the high summand is $Qt+B$, then $B$ plus the low summand is at most $3B<Q$; again there is no carry into the quotient, and necessarily $t=i$ with low summand $j$. These are precisely the two displayed decompositions. The same quotient--remainder description shows $e_{ij}\notin\mathcal V$.
\end{proof}

\subsection{A globally valid exhaustive-merge construction}
A valid addition chain for the desired lengths would not suffice here: in bottom-up tokenisation the rules must actually form the required adjacent pairs in every string simultaneously. The following construction establishes that stronger property.

\begin{lemma}[Binary assembly]\label{unaryope:lem:binary}
Start with the descending binary expansion of each integer $x\le M$, represented as a list of its distinct power-of-two summands. For each non-power-of-two $z=2,\ldots,M$, in increasing order, apply the rule
 $\displaystyle (b,z-b)\longrightarrow z,\qquad b=2^{\lfloor\log_2z\rfloor}.$ 
Every such list becomes its single token $x$. The statement also holds after multiplying all token lengths and rule operands by a common positive integer.
\end{lemma}
\begin{proof}
For a list with highest binary summand $b$, its suffix represents $s=x-b<b$. By induction on the highest binary summand, that suffix is assembled into $s$ using rules with output at most $s$. No non-power rule involving $b$ is applied before output $b+1$, so the suffix has finished before $b$ can participate in any rule. If $s>0$, the list is now $[b,s]$. Among rules with first operand $b$, only the rule with output $b+s=x$ can act on this pair. It makes the whole list a singleton. If $s=0$, the list was a singleton already. Scaling every length preserves all pair matches.
\end{proof}

\begin{lemma}[Base preamble]\label{unaryope:lem:preamble}
There is a sequence $P$ of exactly $h$ exhaustive left-to-right merges, creating precisely $F\setminus\{1\}$, that produces
\begin{align*}
a^f&\longmapsto[f] && (f\in F),\\
a^{\ell_i}&\longmapsto[Qi,B],\\
a^{\rho_j}&\longmapsto[B,j],\\
a^{e_{ij}}&\longmapsto[Qi,B,j].
\end{align*}
\end{lemma}
\begin{proof}
First perform all doublings $(x,x)\to2x$, starting at $x=1$ and ending at the largest power of two at most $pQ$. All their outputs belong to $F$: those below $B$ lie in $\{1,\ldots,r\}$, followed by $B,2B$, and the powers that are multiples of $Q$.

Successive left-to-right doubling gives the descending binary expansion of any input length. Every relevant length is less than $Q(p+1)$. Since $Q$ is a power of two, any power of two greater than $pQ$ is at least $Q(p+1)$; therefore the doublings performed suffice for all relevant strings.

Next, for each non-power-of-two $z\le r$, in increasing order, perform
 $\displaystyle (2^{\lfloor\log_2z\rfloor},\ z-2^{\lfloor\log_2z\rfloor})\to z.$ 
By Lemma~\ref{unaryope:lem:binary}, this assembles each right-hand binary block $j\le r$. It cannot cross the preceding token $B$, which exceeds all operands in this stage.

Finally, for each non-power-of-two $i\le p$, in increasing order, perform
 $\displaystyle (Qb,\ Q(i-b))\to Qi,\qquad b=2^{\lfloor\log_2i\rfloor}.$ 
The scaled form of Lemma~\ref{unaryope:lem:binary} assembles the left block $Qi$. These rules use only multiples of $Q$, so they cannot cross $B$ or interact with a right block $j<Q$.

The initial binary expansions have exactly the separated blocks just described, because $j<B$ and $Q=4B$. This proves all asserted parses. Every non-character base length is created exactly once: powers of two in the first stage, non-power small lengths in the second, and non-power multiples of $Q$ in the third. No other length is created. The total number of rules is therefore $h$.
\end{proof}

Fix any set $S$ of exactly $k$ vertices. After $P$, append the rules
 $\displaystyle (Qi,B)\to\ell_i\quad\text{for }u_i\in S, \qquad (B,j)\to\rho_j\quad\text{for }v_j\in S.$ 
They can be appended in any order. Each rule has a new output length, and is productive on its own vertex anchor.

Base anchors remain singletons. A selected vertex anchor becomes one token; an unselected anchor remains two. An edge starts at $[Qi,B,j]$. If at least one endpoint is selected, the first applicable endpoint rule reduces it to two tokens. When both endpoints are selected, the other rule cannot subsequently apply, because the middle token $B$ has been consumed. If neither endpoint is selected, the edge remains three tokens.

Thus the sequence has exactly $K=h+k$ rules and achieves
\begin{equation}\label{unaryope:eq:upper}
Ah+C\bigl(k+2(n-k)\bigr)+2\cov(S)+3(m-\cov(S))
=Ah+C(2n-k)+3m-\cov(S).
\end{equation}
In particular, writing
 $\displaystyle H=Ah+C(2n-k)+3m,$ 
we have $L^*_{\BU}\le H-\cov_k(G)\le H$.

\subsection{The lower bound, even for unrestricted direct vocabularies}
Consider an optimal direct vocabulary $W$ containing the character $1$ and at most $K$ additional words. By~\eqref{unaryope:eq:sandwich} and the preceding construction, its total cost is at most $H$.

\subsection{Every base word is forced}
If some $f\in F\setminus\{1\}$ is absent from $W$, all $A$ copies of its anchor need at least two tokens. Each of the other $h-1$ anchor types needs at least one. The base anchors alone then cost at least
 $\displaystyle Ah+A>Ah+2nC+3m\ge H,$ 
a contradiction. Hence $F\subseteq W$, and the base contribution is exactly $Ah$.

\subsection{Exactly \texorpdfstring{$k$}{k} vertex words are forced}
The base has already used $h$ of the additional-word budget, so at most $k$ choice lengths can be in $W$. Let $q$ be their number. If $q\le k-1$, the vertex anchors cost at least
 $\displaystyle C\bigl(q+2(n-q)\bigr)\ge C(2n-k+1).$ 
Even allowing every edge string to become a singleton, the overall cost would be at least
 $\displaystyle Ah+C(2n-k+1)+m =H+(C-2m)=H+1,$ 
another contradiction. Thus $q=k$.

These $h+k$ distinct non-character words exhaust the whole budget. Therefore
 $\displaystyle W=F\cup\{\ell_i:u_i\in S\}\cup\{\rho_j:v_j\in S\}$ 
for a set $S$ of exactly $k$ vertices. In particular, no spare helper length, edge length, or duplicate output can bypass the intended choices. This normal form was derived from an arbitrary optimum, not assumed.

\subsection{Edge savings are bounded by coverage}
Every edge needs at least two tokens because its full length is not in $W$. By Lemma~\ref{unaryope:lem:arithmetic}, an edge with neither endpoint in $S$ has no two-token representation at all, so it needs at least three. The resulting lower bound is
 $\displaystyle L^*_{\DIR}\ge Ah+C(2n-k)+3m-\cov(S) \ge H-\cov_k(G).$ 
Combining this with~\eqref{unaryope:eq:upper} and~\eqref{unaryope:eq:sandwich} proves~\eqref{unaryope:eq:identity}.

\subsection{Complexity consequences and scope}
By~\eqref{unaryope:eq:identity} and the definition of $\delta$,
 $\displaystyle L^*_{\BU}(D,K)\le\delta \iff \cov_k(G)\ge t \iff L^*_{\OPE}(D,K)\le\delta.$ 
This proves the required reductions.

\subsection{Polynomial explicit size}
Since $r<B\le2r$, we have $Q\le8r$. Every string length is less than $Q(p+1)=O(n^2)$. Also $h=O(n)$, $C=O(m+1)$, and $A=O(n(m+1))$. There are
 $\displaystyle Ah+Cn+m=O(n^2(m+1))$ 
explicit dataset strings, so their total character count is $O(n^4(m+1))$, at most $O(n^6)$ for a simple graph. The budget $K=n+1+k\le2n+1$ and the threshold are polynomially bounded as well. There is no exponential-length unary string hidden in a binary encoding.

\subsection{Membership in NP}
Let $N$ be the total explicit character count and $M$ the maximum input-string length. For bottom-up tokenisation, no-op rules may be deleted: any token that never appears cannot affect a later parse. Every remaining rule reduces the total token count by at least one, so a certificate requires at most $N$ productive rules. Its operands are words of length at most $M\le N$, and its action can be simulated in polynomial time.

For the OPE-type model, only words of length at most $M$ can be useful. All ancestors of a useful generated word also have length at most $M$, since concatenation adds positive lengths. Keep the relevant generation dependencies, removing duplicate words and unused rules. This leaves at most $M-1$ additional unary words. Alternatively, if $K\ge M-1$, all lengths $1,\ldots,M$ can be generated with $M-1$ additions, making every input a singleton. For a guessed vocabulary $W$, optimal segmentation is verified by the recurrence
 $\displaystyle d[0]=0,\qquad d[x]=1+\min_{w\in W,\,w\le x}d[x-w] \quad(1\le x\le M).$ 
Summing $d[x]$ over all dataset entries verifies the target cost in polynomial time. This establishes NP membership and completes the proof of Theorem~\ref{unaryope:thm:main}.

\begin{corollary}
Unless $\mathsf P=\mathsf{NP}$, neither optimisation problem has a deterministic fully polynomial-time approximation scheme on explicitly represented unary datasets.
\end{corollary}
\begin{proof}
For a nonempty dataset, the optimum is an integer at most its total explicit length $N$. An FPTAS with $\varepsilon=1/(N+1)$ would return a feasible integer cost less than $\OPT+1$, hence exactly $\OPT$, in polynomial time. This contradicts Theorem~\ref{unaryope:thm:main} unless $\mathsf P=\mathsf{NP}$.
\end{proof}

The reduction does not establish APX-hardness or exclude a PTAS: the forcing terms may dominate the variable coverage contribution. It concerns datasets with repetitions, not one unary input string or a dataset with at most two distinct lengths. It does not establish approximation guarantees for unary OPE or standard bottom-up token count.

\section{Conclusions and Open Problems}\label{sec:conclusions}
\subsection{Conclusions}
We have presented a joint account of optimization and hardness results for direct and unary merge-based tokenisation. The main contrast is that binary direct token count admits no $M^{1-\varepsilon}$-approximation under $\mathrm P\ne\mathrm{NP}$ while admitting a polynomial-time $O(M/\log M)$ approximation; maximizing the corresponding savings admits a fixed-alphabet EPTAS. Structural properties of optimal dictionaries further yield subexponential exact algorithms. The APX-hardness of unary bottom-up tokenisation and the strong NP-completeness of unary OPE show that the exact semantics of merges matter.

\subsection{Open problems}\label{sec:open}
Key questions include whether binary direct tokenisation is fixed-parameter tractable in vocabulary budget $K$ alone; whether binary standard bottom-up token count admits any constant-factor approximation; whether unary OPE admits a PTAS; and how closely the polynomial-time bounds for binary direct count can be matched without invoking ETH. Further work should distinguish exact global vocabulary selection from greedy merge-frequency heuristics and should investigate which of the structural exact algorithms extend to alternative parsing semantics.

\medskip
\noindent\textbf{AI disclosure.} The authors used GPT-5.6 Sol during the preparation of this
work to brainstorm and refine constructions, explore proof
strategies, and discuss and check mathematical arguments.
The model also assisted with proofreading and aspects of
the literature review. All mathematical claims, proofs,
and cited results were independently verified by the authors,
who take full responsibility for the accuracy and content
of the paper.
\bibliographystyle{unsrtnat}
\bibliography{references}
\end{document}